\documentclass[
 amsmath,amssymb,
 aps,
 pra,
]{revtex4-2}

\usepackage{braket} %狄拉克符号包

\usepackage{graphicx}% Include figure files
\usepackage{dcolumn}% Align table columns on decimal point
\usepackage{bm}% bold math
\usepackage{hyperref}
\hypersetup{
	colorlinks=true,
	linkcolor=cyan,
	filecolor=blue,
	urlcolor=red,
	citecolor=blue,
}
\begin{document}

\preprint{APS/123-QED}

\title{Residual-Squeezing Mechanism of Mismatch in Inverse-Squeezing Kennedy Receivers}
% \thanks{A footnote to the article title}%

\author{Enhao Bai}
  \affiliation{Information Support Force Engineering University, Wuhan 430035, China}
  \affiliation{Guangxi Key Laboratory of Multimedia Communications and Network Technology, Guangxi University, Nanning 530006, China}

\author{Fengkai Sun}
  \affiliation{Guangxi Key Laboratory of Multimedia Communications and Network Technology, Guangxi University, Nanning 530006, China}

\author{Tianyi Wu}
  % \email{18142630162@163.com}
  \affiliation{Information Support Force Engineering University, Wuhan 430035, China}
\author{Yang Ran}
  \affiliation{Information Support Force Engineering University, Wuhan 430035, China}
\author{Zichao Zhou}
  \affiliation{Information Support Force Engineering University, Wuhan 430035, China}
\author{Huankai Zhang}
  \affiliation{Information Support Force Engineering University, Wuhan 430035, China}
\author{Jian Peng}%
  \email{pengjian@nudt.edu.cn}
  \affiliation{Information Support Force Engineering University, Wuhan 430035, China}
\author{Chen Dong}
  \email{dongchengfkd@163.com}
  \affiliation{Information Support Force Engineering University, Wuhan 430035, China}

\author{Laiyuan Tong}
  \affiliation{Guangxi Key Laboratory of Multimedia Communications and Network Technology, Guangxi University, Nanning 530006, China}
\author{Zhenrong Zhang}
  \email{zzr76@gxu.edu.cn}
  \affiliation{Guangxi Key Laboratory of Multimedia Communications and Network Technology, Guangxi University, Nanning 530006, China}

\author{Yaping Li}
  \affiliation{Wuhan Maritime Communication Research Institute, Wuhan 430035, China}

\date{\today}

\begin{abstract}
The discrimination of quantum states is fundamental to quantum information processing. Inverse-squeezing Kennedy (IS-Kennedy) receivers can outperform the coherent-state BPSK Helstrom benchmark at the same energy by converting transmitter-side squeezing into an effective coherent-state separation gain, without violating the Helstrom bound for the squeezed-state alphabet. This work investigates how squeezing mismatch degrades this mechanism. We show that imperfect inverse squeezing transforms the ideally nulled output into a residually squeezed state, thereby altering the photon-number statistics before detection. This residual-squeezing picture reveals a strong physical asymmetry between squeezing-magnitude and squeezing-phase mismatches. Magnitude mismatch produces an energy-independent error floor in the high-signal-energy regime, whereas phase mismatch generates a residual squeezing term that grows with signal energy. In the small-residual-squeezing regime, this leads to a polynomial growth of the leading error contribution and a rapid collapse of the SQL advantage. We also identify a parity-step effect in photon-number-resolving detection: because the nulled residual squeezed vacuum contains only even photon numbers, increasing detector resolution improves the high-energy robustness only when the effective saturation threshold crosses the next even photon number. These results identify phase locking as the dominant bottleneck for IS-Kennedy-type non-Gaussian receivers under unitary squeezing mismatch and provide design guidelines for robust squeezed-state quantum receivers.

\end{abstract}

% \keywords{Squeezing mismatch, photon-number-resolving detector, quantum receiver}

\maketitle

%\tableofcontents

\section{Introduction}
Quantum state discrimination is a central task in quantum communication and quantum measurement theory. 
For binary optical communication, classical measurements such as homodyne detection or direct detection are generally limited by the standard quantum limit (SQL) \cite{review_2021}, and the minimum error probability is set by the Helstrom bound (HB) which is governed by the non-orthogonality of quantum states \cite{Helstrom}.
A major objective in quantum receiver design is therefore to construct physically implementable measurements that approach the Helstrom bound while using experimentally accessible optical operations. 
Prominent examples include static displacement receivers \cite{Kennedy, Optimal_Displacement, zhaomufei_2020}, adaptive displacement receivers \cite{Dolinar, CPN, Partition_Adaptive_Nulling_Receiver}, classical-quantum hybrid receivers \cite{Hybrid_Receiver, Olivares_01, Bai_1st}, and Gaussian-operation-assisted receivers \cite{shuro_squeeze, sun_bpsk_multichannel}.

Most studies of quantum receivers have focused on coherent-state alphabets, because of the experimental accessibility and robustness under bosonic loss of coherent state \cite{coherent_state_01, coherent_state_02, coherent_state_03}.
A different route is to exploit nonclassical transmitter states whose intrinsic overlap is smaller than that of coherent states at the same energy \cite{Paris_2001, Olivares2018, DSS_2025_arxiv}.
Reference~\cite{Olivares2018} demonstrates that, under the same energy, the Helstrom bound for binary phase-shift-keyed (BPSK) displaced squeezed vacuum states (S-BPSK) is lower than that for coherent states (C-BPSK). 
An inverse-squeezing Kennedy (IS-Kennedy) receiver has recently been introduced for S-BPSK discrimination \cite{IS_Kennedy}. In the ideal case, the receiver first applies a Kennedy nulling displacement to map one hypothesis to a squeezed vacuum and the other to a displaced squeezed state. It then applies an inverse-squeezing (IS) operation matched to the signal squeezing. When this matching is exact, the squeezed on-off-keyed alphabet is transformed into a coherent-state on-off-keyed alphabet. As a result, the squeezing resource is converted into an effective coherent-energy gain before photon counting, allowing the receiver to outperform the S-BPSK SQL and the C-BPSK HB over a broad range of signal energies.
In Ref.~\cite{IS_Kennedy}, we further investigated the IS-Kennedy receiver under several practical imperfections, including channel phase diffusion and pure transmission loss. The results showed that the receiver can retain a discrimination advantage under phase diffusion over finite signal-energy ranges. Under transmission loss, although the squeezing-enabled gain is progressively suppressed with increasing propagation loss, the IS-Kennedy can still outperform the conventional Kennedy receiver over a finite range of link lengths, particularly in the low-signal-energy regime.

This ideal mechanism, however, relies on an exact cancellation between the signal squeezing and the receiver squeezing.
In practice, the inverse-squeezing operation is affected by finite calibration accuracy, pump fluctuations, and phase-locking errors \cite{IS_mismatch_01,IS_mismatch_02}.
In this work, we investigate mismatch in the IS-Kennedy receiver for S-BPSK discrimination. We show that mismatch is equivalent to a residual squeezing before photon-number detection, which makes the optimal maximum-a-\emph{posteriori} (MAP) decision generally non-single-threshold. 
We further show that the receiver is much more sensitive to phase mismatch than to magnitude mismatch. 
These results clarify the physical origin of mismatch-induced degradation and identify the phase-locking accuracy required for practical implementations of inverse-squeezing receivers.

The rest of this work is organized as follows. 
Section II introduces the ideal inverse-squeezing Kennedy receiver.
Section III formulates the mismatch model. 
Section IV presents numerical simulations and mismatch-tolerance maps relative to the S-BPSK SQL. 
Section V summarizes the key findings and analyzes potential application scenarios.

\section{Modeling the ideal IS-Kennedy receiver}
The S-BPSK signals to be discriminated are given by \cite{Olivares2018}
\begin{equation}
  \begin{aligned}
    &\text{hypothesis}\ '0': \ket{\psi_0} = D(-\alpha)S(r)\ket{0} \sim \rho_0,\\
    &\text{hypothesis}\ '1': \ket{\psi_1} = D(+\alpha)S(r)\ket{0} \sim \rho_1.
  \end{aligned}
\end{equation}
Without loss of generality, $\alpha$ is chosen to be real, and the mean photon numbers of the signals are $N=|\alpha|^2 + \sinh^2(r)$.
Throughout this work, equal a priori probabilities are assumed, i.e., $\pi_0 = \pi_1 = 0.5$.

Reference \cite{Olivares2018} defines the squeezing fraction $\beta$ (the share of the total energy stored in squeezing) to quantify the advantage of S-BPSK:
\begin{equation}
  \beta \triangleq \frac{\sinh^2(r)}{N}.
\end{equation}
In addition, under ideal condition, the optimal squeezing fraction is $\beta_\text{opt}=\frac{N}{2N+1}$, at which point Helstrom bound and SQL of S-BPSK are minimized \cite{Olivares2018}. In this work, all S-BPSK employ the optimal squeezing fraction $\beta_\text{opt}$:
\begin{equation}
  \alpha = \sqrt{N(1-\beta_\text{opt})},\quad r = \sinh^{-1}\sqrt{N\beta_\text{opt}},\quad \alpha \text{e}^r = \sqrt{N(N+1)}
\end{equation}

\begin{figure}[htbp]
  \centering
  \includegraphics[scale=1.0]{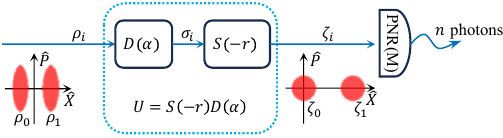}
  \caption{Schematic of the inverse-squeezing Kennedy (IS-Kennedy) receiver for binary displaced squeezed-state discrimination.
  The input state $\rho_i$ is first displaced by $D(\alpha)$ to perform Kennedy nulling, producing the intermediate squeezed on-off-keyed state $\sigma_i$.
  An inverse-squeezing operation is then applied, followed by a photon-number-resolving detector with resolution M, denoted PNR(M).
  The detected photon number is finally processed using the MAP decision rule.}
  \label{fig:structure_of_IS-Kennedy}
\end{figure}

We model the inverse-squeezing Kennedy receiver \cite{IS_Kennedy} as follows. Figure~\ref{fig:structure_of_IS-Kennedy} schematically illustrates the receiver.

Under ideal condition, firstly, the received signal interferes with a local oscillator (LO) on a highly transmissive beam splitter ($\tau \to 1$) to perform the Kennedy nulling displacement $D(\alpha)$. Consequently, the S-BPSK alphabet is mapped to a squeezed-state on-off-keying (S-OOK) alphabet $\sigma_i$:
\begin{equation}
  \sigma_i \sim D(\alpha)\ket{\psi_i} = \left\{\begin{aligned}
    &S(r)\ket{0}, & i = 0,\\
    &D(2\alpha)S(r)\ket{0}, & i = 1.
  \end{aligned}\right.
  \label{eq:ISK_sigma_compact}
\end{equation}
Subsequently, $\sigma_i$ passes through the IS module $S(-r)$, mapping the S-OOK alphabet to a coherent-state on-off-keying (C-OOK) alphabet $\zeta_i$:
\begin{equation}
  \zeta_i = S(-r) \sigma_i S^\dagger(-r) \sim \left\{\begin{aligned}
    &S(-r)S(r)\ket{0} = \ket{0}, & i = 0,\\
    &S(-r)D(2\alpha)S(r)\ket{0} = D(2\alpha \mathrm{e}^r)\ket{0} = \ket{2\gamma}. & i = 1.\\
  \end{aligned} \right.
  \label{eq:ISK_zeta_compact}
\end{equation}
In the general case, $\gamma = \alpha \cosh r + \alpha^* \sinh r$. In this work $\alpha \in \mathbb{R}_+$, so $\gamma = \alpha e^r$. Therefore, we define:
\begin{equation}
  \gamma \triangleq \alpha e^r,\quad N_\text{eff} \triangleq |\gamma|^2 = |\alpha|^2 e^{2r} = N (N+1).
\end{equation}
Next, PNR detection is performed on $\zeta_i$, yielding an output photon number $n$. The positive operator-valued measure (POVM) for an ideal PNR detector with resolution M, denoted as PNR(M), is given by:
\begin{equation}
  \Pi^{(\text{M})}_n = \left\{\begin{aligned}
    &\Pi_n, \quad n =0,1,\cdots, \text{M}-1\\
    &\mathbb{I} - \sum_{k=0}^{\text{M}-1} \Pi_k, \quad n = \text{M}
  \end{aligned}\right.
\end{equation}
where $\Pi_n = \ket{n}\bra{n}, \ n = 0,1,2,\cdots$ and $\mathbb{I}$ is the identity matrix in the Fock representation.

Therefore, the  probability of detecting $n$ photons given hypothesis $'i'$ is given by
\begin{equation}
  P(n|i) = \text{Tr}\left\{\zeta_i\cdot\Pi_n^{(\text{M})}\right\} = \left\{\begin{aligned}
    &\text{Poiss}(n;\mu_i),\quad n =0,1,\cdots, \text{M}-1\\
    &1 - \sum_{k=0}^{\text{M}-1}\text{Poiss}(k;\mu_i),\quad n = \text{M}
  \end{aligned}\right.
  \label{eq:zeta_n_prob}
\end{equation}
where $\text{Poiss}(n;\mu) = \text{e}^{-\mu}\frac{\mu^n}{n!}$; $\mu_0 = 0$ and $\mu_1 = 4N_\text{eff}$ are the mean photon number of $\zeta_0$ and $\zeta_1$, respectively. Then, the \emph{posterior} probabilities are given by:
\begin{equation}
  P(i|n) = \frac{\pi_i P(n|i)}{\sum_{k} \pi_k P(n|k)} = \frac{P(n|i)}{\sum_{k} P(n|k)}
\end{equation}
The MAP decision rule is given by:
\begin{equation}
  \hat{i}(n) = \left\{\begin{aligned}
    &1,\quad P(i = 1|n) \ge P(i = 0|n),\\
    &0,\quad \text{otherwise}.
  \end{aligned}\right.
\end{equation}
where $\hat{i}$ is the decision of IS-Kennedy receiver.
When the likelihood function $L(n) = \ln\frac{P(n|i=0)}{P(n|i=1)}$ is monotonic, the MAP decision is equivalent to a single-threshold decision\cite{IS_Kennedy}, with the optimal threshold denoted as $n_\text{th}^*$. For the exact-nulling configuration considered here, under ideal inverse-squeezing conditions, the optimal threshold is $n_\text{th}^*=1$, which means that PNR detector does not provide a performance advantage over single photon detector (SPD) in this case \cite{IS_Kennedy}.
So the error probability of IS-Kennedy is given by
\begin{equation}
  \begin{aligned}
    P_\text{err}^\text{ISK, ideal} &= \frac{1}{2}\left(P_\text{FA} + P_\text{Mi}\right) = \frac{1}{2}\left[P(n\ge n_\text{th}^*|i=0) + P(n< n_\text{th}^*|i=1)\right]\\
    &= P_\text{err}^\text{K,ideal}(N_\text{eff}) = \frac{1}{2}\exp\left[-4N(N+1)\right],
  \end{aligned}
  \label{eq:isk_err_ideal}
\end{equation}
where $P_\text{FA}$ and $P_\text{Mi}$ are the false alarm probability (deciding 1 when 0 is sent) and the miss probability (deciding 0 when 1 is sent), respectively, and $P_\text{err}^\text{K,ideal}$ is the ideal probability of the Kennedy receiver \cite{Kennedy}. From this, it can be seen that the IS-Kennedy converts the squeezing resources of S-BPSK signals into higher coherent energy advantage ($N\to N_\text{eff}=N(N+1)$) through the IS module, thereby achieving a lower error probability. Compared to quantum receiver for coherent state discrimination, the energy gain brought by the IS-Kennedy takes the form
\begin{equation}
  G = 10\cdot \log_{10} \left(\frac{N_\text{eff}}{N}\right) = 10\cdot \log_{10} \left(N+1\right)\,(\text{dB})
\end{equation}

\emph{Relation to optimized displacement.} In the present work, we adopt the exact Kennedy-nulling displacement $D(\alpha)$ in order to isolate the effect of inverse-squeezing mismatch. We emphasize, however, that exact nulling is not in general the optimal displacement for minimum-error discrimination, as is well known for generalized displacement receivers \cite{Optimal_Displacement}. To see the connection explicitly, consider replacing $D(\alpha)$ by $D(\alpha+\delta)$, where $\delta$ denotes an additional displacement offset. Under ideal inverse squeezing,
\begin{equation}
  S(-r)D(\alpha + \delta) \ket{\psi_0} = \ket{\delta_\text{eff}},\quad
  S(-r)D(\alpha + \delta) \ket{\psi_1} = \ket{2\gamma + \delta_\text{eff}},
\end{equation}
where $\delta_\text{eff} = \delta \cosh r + \delta^* \sinh r$. Thus, optimizing the displacement in an ideally matched IS-Kennedy receiver is mathematically equivalent to the optimized-displacement problem for a coherent-state receiver, with the coherent amplitudes rescaled by the inverse-squeezing gain. Consequently, $\delta=0$, corresponding to exact nulling, is not expected to be universally optimal. We nevertheless retain the exact-nulling configuration throughout this work because our objective is to isolate and characterize the degradation mechanism caused specifically by squeezing mismatch.

\section{Modeling the IS-Kennedy under squeezing mismatch condition}
In practical experimental setups, deviations between the receiver’s squeezing parameters and the input signal are inevitable due to factors such as finite calibration accuracy, pump fluctuations, and phase-locking errors. This phenomenon is referred to as squeezing mismatch. Without loss of generality, we adopt a simplified model where the transmitter performs perfect squeezing and the optical path is lossless, attributing all squeezing adaptation errors solely to the IS module. We denote the squeezing parameter applied by the IS module by
\begin{equation}
  z_s = \left(-r+\Delta r\right)\mathrm{e}^{j\Delta \theta},\quad j = \sqrt{-1},
\end{equation}
where $\Delta r$ represents the magnitude mismatch and $\Delta \theta$ represents the phase mismatch. Under ideal conditions, $\Delta r = \Delta \theta = 0$.

Under mismatch conditions, the overall transformation of the receiver is given by:
\begin{equation}
  \label{eq:zeta_sm}
  \zeta_i^{\text{sm}} \sim S\left(z_s\right) D(\alpha) \ket{\psi_i} = \left\{\begin{aligned}
    &S_\text{tot} \ket{0}, & i = 0,\\
    &S_\text{tot} \ket{2\gamma}, & i = 1.
  \end{aligned}\right.
\end{equation}
Here, the composite squeezing operator is defined as $S_\text{tot} \triangleq S\left(z_s\right) S(r)$, which reduces to the identity operator $S_\text{tot} = \mathbb{I}$ in the ideal case. Thus, the essence of the mismatch is encapsulated within $S_\text{tot}$.

More generally, if an additional displacement offset $\delta$ is introduced, Eq.~\eqref{eq:zeta_sm} becomes
\begin{equation}
  \zeta_0^{\text{sm}} \sim S_\text{tot} \ket{\delta_\text{eff}},\quad
  \zeta_1^{\text{sm}} \sim S_\text{tot} \ket{2\gamma + \delta_\text{eff}}.
\end{equation}
Therefore, in the presence of squeezing mismatch, displacement optimization becomes coupled to the residual-squeezing parameters. Jointly optimizing the displacement and inverse-squeezing settings may further reduce the discrimination error, but such an optimization would introduce an additional control dimension beyond the residual-squeezing mechanism studied here and is left for future work.

\begin{figure*}[htbp]
  \centering
  \includegraphics[scale=1.0]{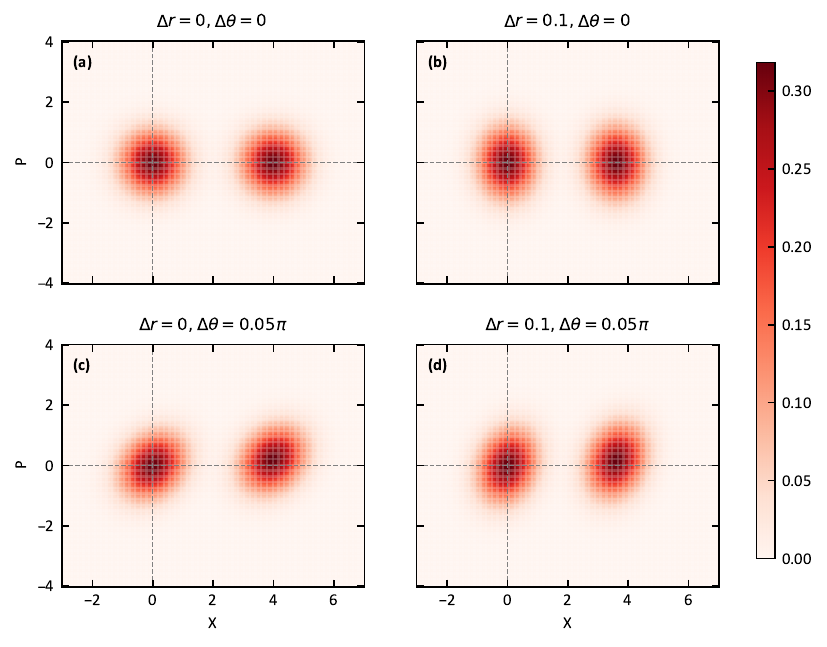}
  \caption{The Wigner functions of $\zeta_0^\text{sm}$ and $\zeta_1^\text{sm}$ with mean photon number of $N = 1.0$, under different squeezing-mismatch cases.}
  \label{fig:phase_space_changes}
\end{figure*}

Using the single-mode Gaussian operator decomposition, this operator can always be expressed as:
\begin{equation}
  S_\text{tot} = R(\vartheta)\,S(z_m),
  \label{eq:Stot_decomp}
\end{equation}
where $R(\vartheta)=\exp(-j\vartheta a^\dagger a)$ is a phase rotation, and $S(z_m)$ represents the equivalent residual squeezing with parameter $z_m=r_m e^{j\theta_m}$. Since $R(\vartheta)$ contributes only a phase factor in the Fock basis and does not alter the photon number distribution, the impact of mismatch on PNR statistics for a given energy $N$ is determined solely by the residual squeezing parameters $(r_m,\theta_m)$.

Next, we determine $(r_m, \theta_m)$ using the Bogoliubov transformation. Letting $r_s \triangleq -r+\Delta r$ and $\theta_s \triangleq \Delta\theta$, the coefficients for the transformation $S^\dagger_\text{tot} a S_\text{tot} = x a + y a^\dagger$ are derived as:
\begin{equation}
  x = \cosh(r_s)\cosh r + \mathrm{e}^{j\theta_s}\sinh(r_s)\sinh r,\qquad
  y = -\cosh(r_s)\sinh r - \mathrm{e}^{j\theta_s}\sinh(r_s)\cosh r,
\end{equation}
satisfying the unitarity condition $|x|^2-|y|^2=1$. Consequently, we obtain:
\begin{equation}
  r_m=\mathrm{arcsinh}(|y|),\qquad
  \vartheta=-\arg(x),\qquad
  \theta_m=\arg(-y)-\arg(x).
  \label{eq:rm_theta_from_munu}
\end{equation}

In practical scenarios characterized by small mismatches ($\Delta r\ll 1,\ |\Delta\theta|\ll 1$)~\cite{IS_mismatch_01,IS_mismatch_02}, a first-order approximation yields:
\begin{equation}
  y \approx -\Delta r + j\,\frac{\Delta\theta}{2}\sinh(2r),
  \label{eq:nu_approx}
\end{equation}
which leads to:
\begin{equation}
  \begin{aligned}
    &r_m \approx \sqrt{(\Delta r)^2+\left(\frac{\Delta\theta}{2}\sinh 2r\right)^2 }, \\
    &\theta_m \approx \arg \left(\Delta r - j \frac{\Delta\theta}{2} \sinh\left(2r\right)\right) + \Delta \theta \sinh^2 r.
  \end{aligned}
  \label{eq:rm_approx}
\end{equation}
Equation~\eqref{eq:rm_approx} reveals that the total residual squeezing $r_m$ is significantly more sensitive to the phase mismatch $\Delta\theta$ than to the magnitude mismatch $\Delta r$. This arises because the phase mismatch term is scaled by the coefficient $\frac{1}{2}\sinh(2r)$, which grows rapidly with the squeezing parameter $r$ (and thus the signal energy $N$).
This leads to a ``dominant masking effect'': in the presence of non-negligible phase mismatch, its contribution to $r_m$ typically far exceeds that of the magnitude mismatch. Given the geometric relationship $r_m \approx \sqrt{(\Delta r)^2+\left(K\Delta\theta\right)^2 }$, the value of $r_m$ becomes almost entirely determined by the phase mismatch term. This explains the phenomenon observed in the simulation results, as illustrated in Fig.~\ref{fig:performance_squeezing_mismatch}(a) and (b) in Sec.~IV, where the error probability curves for $(\Delta r, \Delta \theta) = (0.00, 0.03\pi)$ and $(0.02, 0.03\pi)$ nearly coincide. Consequently, in experimental systems, the requirement for phase locking precision is much more stringent than that for magnitude matching.

For phase-only mismatch ($\Delta r=0$), we obtain
\begin{equation}
  \label{eq:phase_mis_parameter}
  \begin{aligned}
    &r_m \approx \frac{|\Delta\theta|}{2}\sinh(2r)  =|\Delta\theta|\,\frac{N(N+1)}{2N+1},\\
    &\theta_m \approx -\frac{\pi}{2}\,\mathrm{sgn}(\Delta\theta)  +\Delta\theta\,\sinh^2 r  =-\frac{\pi}{2}\,\mathrm{sgn}(\Delta\theta)+\Delta\theta\,\frac{N^2}{2N+1},\ (\mathrm{mod}\ 2\pi).
  \end{aligned}
\end{equation}
For magnitude-only mismatch ($\Delta\theta=0$), we obtain
\begin{equation}
  r_m \approx |\Delta r|,\qquad
  \theta_m \approx
  \begin{cases}
    0, & \Delta r>0,\\
    \pi, & \Delta r<0,
  \end{cases}
  \quad (\mathrm{mod}\ 2\pi).
\end{equation}

We now calculate the photon number probability distribution of $\zeta_i^{\text{sm}}$. Ignoring the rotation $R(\vartheta)$ which does not affect photon statistics, the equivalent output states derived from Eq.~\eqref{eq:Stot_decomp} are:
\begin{equation}
  \zeta_0^\text{sm} \sim S(z_m)\ket{0},\qquad
  \zeta_1^\text{sm} \sim S(z_m)\ket{2\gamma}.
\end{equation}
To utilize standard photon number distribution formulas, we rewrite $\zeta_1^\text{sm}$ as a displaced squeezed vacuum state:
\begin{equation}
  \zeta_1^\text{sm} \sim S(z_m)\ket{2\gamma} = D(2\gamma_m)\,S(z_m)\ket{0},
  \label{eq:sc_to_dss}
\end{equation}
where
\begin{equation}
  \gamma_m=\gamma\cosh r_m - \gamma^\ast \mathrm{e}^{i\theta_m}\sinh r_m.
  \label{eq:gamma_m_def}
\end{equation}
Consequently, the Poissonian distribution valid for the ideal IS-Kennedy (Eq.~\eqref{eq:zeta_n_prob}) no longer applies. For an ideal PNR(M) detector (considering only finite resolution), the probability of detecting $n$ photons given hypothesis $'i'$ can be written as
\begin{equation}
  \label{eq:fock-population_mismatch_squeeze}
  P(n|i)=\text{Tr}\left(\Pi_n\,\zeta_i^{\text{sm}}\right)=
  \left\{\begin{aligned}
    &p_n^{\rm svs}(r_m), & i=0,\\
    &p_n^{\rm dss}(2\gamma_m,r_m,\theta_m), & i=1,
  \end{aligned}\right.
\end{equation}
where the corresponding distribution is used for $n \le \text{M}-1$, and the truncation/saturation bin is defined as $P(n=\text{M}|i)=1-\sum_{n=0}^{\text{M}-1}P(n|i)$. Here, $p_n^{\rm svs}(r_m)$ denotes the photon number distribution of the squeezed vacuum state (Eq.~\eqref{eq:pn_svs}), and $p_n^{\rm dss}(2\gamma_m,r_m,\theta_m)$ denotes that of the displaced squeezed vacuum state (Eq.~\eqref{eq:pn_dss}).

The introduction of mismatch results in $\zeta_0^\text{sm}$ becoming a squeezed vacuum state (containing only even photon numbers), while $\zeta_1^\text{sm}$ becomes a displaced squeezed vacuum state whose Fock distribution generally exhibits even-odd oscillations, as shown in Fig.~\ref{fig:performance_squeezing_mismatch}(d). This implies that the likelihood function $L(n)$ does not necessarily vary monotonically with the photon number $n$. Consequently, the MAP decision generally becomes a set-based decision (potentially involving a multi-threshold structure) rather than a simple single-threshold test. Therefore, the minimum error probability (Eq.~\eqref{eq:isk_err_ideal}) is reformulated as:
\begin{equation}
  P_\text{err}^{\text{ISK},\text{sm}} 
  = 1 - \frac{1}{2} \sum_{n=0}^{\text{M}} \max \left\{P(n|i=0),\ P(n|i=1)\right\}.
\end{equation}

When employing an SPD, the observation yields only two outcomes: $n=0$ (no-click) and $n \ge 1$ (click). Under equal priors, the MAP decision is equivalent to a binary mapping of these outcomes. Since $P(0|0) \ge P(0|1)$, the MAP rule maps to on--off detection: click decides hypothesis $'1'$, no-click decides hypothesis $'0'$. The false alarm probability is given by
\begin{equation}
  P_{\rm FA}=P(\text{click}|0)=1-P(n=0|i=0)=1-\frac{1}{\cosh r_m}.
\end{equation}
The miss detection probability is given by
\begin{equation}
  P_{\rm Mi}=P(n=0|i=1)=\frac{1}{\cosh r_m}\exp\left(
    -|2\gamma_m|^2+\mathrm{Re}\big[\mathrm{e}^{-i\theta_m}(2\gamma_m)^2\big]\tanh r_m
  \right).
\end{equation}
These equations clearly reveal that the residual squeezing $r_m$ reduces the vacuum component of $\zeta_0^\text{sm}$, thereby increasing the false alarm probability, while $\theta_m$ modifies the vacuum overlap of $\zeta_1^\text{sm}$ through the exponential term.

To gain deeper insight into the distinct behaviors of magnitude and phase mismatches, we can derive a unified analytical expression for the error probability in the high-signal-energy regime.
As the signal energy $N$ increasing, the mean photon number of the signal state $\zeta_1^{sm}$ far exceeds the resolution $\text{M}$ of the PNR detector, leading to a vanishing miss probability ($P_\text{Mi} \rightarrow 0$).
In this regime, the error probability of the IS-Kennedy receiver is entirely dominated by the false alarm probability.
The MAP strategy is thus equivalent to a threshold decision with $n_\text{th}^* = \text{M}$: decide hypothesis $'1'$ if the outcome falls into the saturation bin $n=\text{M}$, and hypothesis $'0'$ otherwise. 
Since $\zeta_0^\text{sm}$ is a squeezed vacuum state with a non-zero photon number distribution only for even terms, as given by Eq.~\eqref{eq:pn_svs}, the saturation error probability is given by
\begin{equation}
  P_\text{err} \approx \frac{1}{2} P_\text{FA} 
  = \frac{1}{2} \sum_{k = {\left \lceil M/2 \right \rceil }}^{\infty} p_{2k}^{\text{svs}}(r_m)
  = \frac{1}{2} \sum_{k = {\left \lceil M/2 \right \rceil }}^{\infty} \frac{(2k)!}{2^{2k}(k!)^2}\frac{\tanh^{2k}(r_m)}{\cosh(r_m)}.
  \label{eq:sat_both}
\end{equation}
Under the condition of small residual squeezing ($r_m \ll 1$), we can apply the approximations $\tanh(r_m) \approx r_m$ and $\cosh(r_m) \approx 1$. The summation is then dominated by its leading term. Defining $n_{min} = 2\lceil M/2 \rceil$ as the minimum even-photon threshold, we obtain a unified polynomial approximation:
\begin{equation}
  P_\text{err} \approx \frac{1}{2} p_{n_\text{min}}^{\text{svs}}(r_m) 
  = \frac{1}{2} \frac{n_\text{min}!}{2^{n_\text{min}}[(n_\text{min}/2)!]^2} \left(r_m\right)^{n_\text{min}}.
\end{equation}

For the magnitude-only mismatch ($\Delta r \neq 0, \Delta\theta = 0$), the residual squeezing is independent of the signal energy, given by $r_m \approx |\Delta r|$. Substituting this into the unified formula yields:
\begin{equation}
  P_\text{err} \approx \frac{1}{2}\frac{n_\text{min}!}{2^{n_\text{min}}[(n_\text{min}/2)!]^2}(|\Delta r|)^{n_\text{min}}.
  \label{eq:sat_magnitude}
\end{equation}
Since this expression is a constant independent of $N$, the error probability inevitably saturates to a fixed floor.

Conversely, for the phase-only mismatch ($\Delta r = 0, \Delta\theta \neq 0$), the residual squeezing parameter grows monotonically with the signal energy.
Using Eq.~\eqref{eq:phase_mis_parameter}, we have $r_m \approx \frac{|\Delta\theta|}{2}\sinh(2r) = |\Delta\theta|\frac{N(N+1)}{2N+1}$. Substituting this into the general formula gives:
\begin{equation}
  P_\text{err} \approx \frac{1}{2}\frac{n_\text{min}!}{2^{n_\text{min}}[(n_\text{min}/2)!]^2} \left( |\Delta\theta|\frac{N(N+1)}{2N+1} \right)^{n_\text{min}}.
  \label{eq:sat_phase}
\end{equation}
In the large-$N$ limit, the term $\frac{N(N+1)}{2N+1}$ scales asymptotically as $N/2$. Consequently, the error probability scales as $P_\text{err} \propto (|\Delta\theta|N)^{n_\text{min}}$. This reveals a critical physical distinction: rather than producing an $N$-independent floor, phase-only mismatch yields a leading error contribution that grows polynomially with $N$ in the small-residual-squeezing regime.

Finally, a critical observation can be drawn from Eqs.~\eqref{eq:sat_both}, \eqref{eq:sat_magnitude} and \eqref{eq:sat_phase} regarding the role of the detector resolution $\text{M}$.
In the high-signal-energy regime, the error probability is exclusively governed by the exponent $n_{min} = 2\lceil M/2 \rceil$, regardless of the specific mismatch type.
Because the value of $n_\text{min}$ only increments when $\text{M}$ crosses an even integer (for instance, transitioning from $M=2$ to $M=3$), increasing the PNR resolution does not yield a continuous reduction in the high-energy error probability.
Instead, an improvement in the error scaling is only achieved when the resolution is effectively increased in steps of two, namely ``parity-step effect".

\section{Simulation methods and results}
The numerical results were obtained using QuTiP in Python \cite{qutip}. We model the receiver output states in a truncated Fock space and compute the photon-number distributions from the corresponding density operators. Unless otherwise specified, the simulations use a cutoff $N_\text{cut} = 100$. We verified numerical convergence by increasing the cutoff dimension until the resulting error probabilities changed negligibly.
For each mean photon number $N$, the S-BPSK states were generated with the optimal squeezing fraction $\beta_\text{opt}=\frac{N}{2N+1}$. 

\begin{figure*}[htbp]
  \centering
  \includegraphics[scale=1.0]{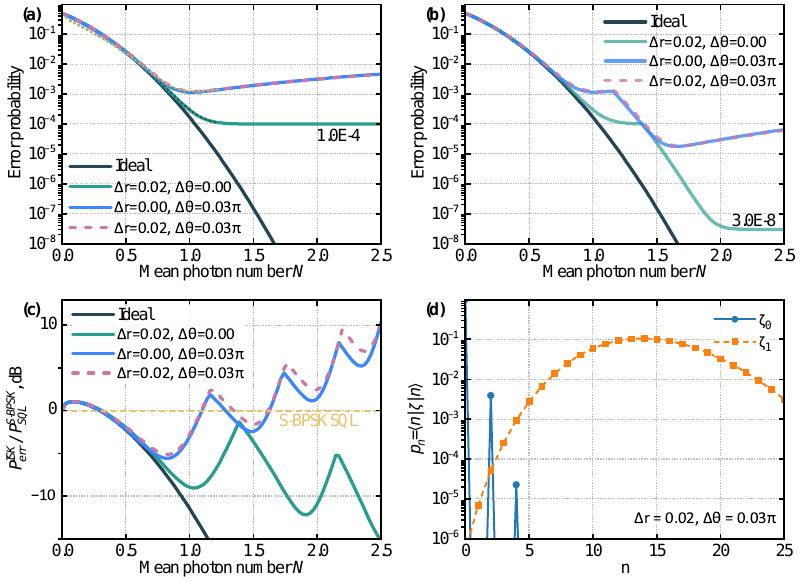}
  \caption{Performance analysis of the IS-Kennedy receiver under inverse-squeezing mismatch conditions. (a) Error probability of IS-Kennedy with SPD. (b) Error probability of IS-Kennedy with PNR(3). (c) The performance ratio (in dB) of IS-Kennedy with PNR(10) relative to the S-BPSK SQL ($P_\text{SQL}^\text{S-BPSK} = \frac{1}{2}\text{erfc}\left(\sqrt{2N(N+1)}\right)$). (d) The Fock-basis population of $\zeta_0^\text{sm},\ \zeta_1^\text{sm}$ with inverse-squeezing mismatch of $\Delta r = 0.02,\ \Delta \theta = 0.03\pi$. }
  \label{fig:performance_squeezing_mismatch}
\end{figure*}

Figures~\ref{fig:performance_squeezing_mismatch}(a) and \ref{fig:performance_squeezing_mismatch}(b) illustrate the error probability for the IS-Kennedy receiver with an SPD (PNR(1)) and a PNR(3) detector, respectively. Several key observations can be made:

First, in the low-signal-energy regime ($N < 0.5$), the error probability curves under various mismatch conditions closely track the ideal curve. This suggests that the IS-Kennedy receiver is naturally robust to inverse-squeezing mismatches at low energies.

Second, there is a striking asymmetry between magnitude and phase mismatches. While magnitude-only mismatch ($\Delta r = 0.02, \Delta\theta = 0$) leads to a performance departure at moderate $N$, its impact is much less severe than that of phase mismatch. Notably, the curve for combined mismatch ($\Delta r = 0.02, \Delta\theta = 0.03\pi$) almost overlaps with the phase-only mismatch curve. This phenomenon confirms the ``dominant masking effect" predicted by Eq.~\eqref{eq:rm_approx}, where the phase error, amplified by the squeezing process, becomes the primary driver of $r_m$ and thus dictates the overall performance degradation.

Third, the high-energy behaviors reveal a fundamental physical distinction between the two mismatches.
The magnitude-only mismatch curve eventually flattens into a constant error floor.
In sharp contrast, curves containing phase mismatch fail to saturate over the plotted range;
after reaching a minimum, the error probability rises with $N$, consistent with the polynomial scaling predicted by Eq.~\eqref{eq:sat_phase} in the small-residual-squeezing regime.
This behavior confirms that magnitude mismatch yields an $N$-independent error floor, whereas phase mismatch leads to an $N$-dependent degradation of the SQL advantage.

Fourth, while increasing the PNR resolution can effectively lower the minimum error probability, this improvement exhibits a characteristic ``parity-step" behavior.
Specifically, enhancing the detector resolution $\text{M}$ only yields a tangible reduction in the error probability when the increment crosses an even-integer threshold.
The physical origin of this phenomenon lies in the state transformation under mismatch: $\zeta_0^\text{sm}$ is a squeezed vacuum state, whose photon-number distribution contains only even photon numbers, as depicted in Fig.~\ref{fig:performance_squeezing_mismatch}(d).

Figure~\ref{fig:performance_squeezing_mismatch}(c) further plots the error-probability ratio in dB relative to the S-BPSK SQL, where negative values indicate performance below the SQL error probability. 
In the ideal case, the IS-Kennedy receiver preserves a clear advantage as $N$ increases.
Under magnitude-only mismatch, this advantage is largely maintained over a broad range of $N$, but is eventually limited by the saturation effect.
By contrast, phase mismatch rapidly reduces the SQL advantage, and the error probability even continues to rise at large $N$.

\begin{figure*}[htbp]
  \centering
  \includegraphics[scale=1.0]{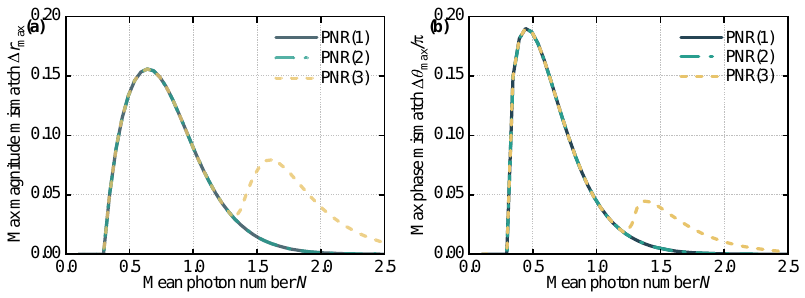}
  \caption{The maximum tolerable inverse-squeezing mismatch for which the IS-Kennedy receiver still outperforms the S-BPSK SQL, i.e., $P_\text{err}^\text{ISK}<P_\text{SQL}^\text{S-BPSK}$. (a) Maximum tolerable magnitude mismatch $\Delta r_\text{max}$ versus mean photon number $N$. (b) Maximum tolerable phase mismatch $\Delta \theta_\text{max} / \pi$ versus $N$. Results are shown for PNR(1), PNR(2), and PNR(3).}
  \label{fig:max_squeezing}
\end{figure*}

Figure~\ref{fig:max_squeezing} provides a more direct robustness metric by showing the maximum tolerable inverse-squeezing mismatch for which the IS-Kennedy receiver still outperforms the S-BPSK SQL, i.e., $P_\text{err}^\text{ISK}<P_\text{SQL}^\text{S-BPSK}$.
As shown in Fig.~\ref{fig:max_squeezing}(a), the tolerable magnitude mismatch $\Delta r_\text{max}$ is non-monotonic in $N$. It first increases,reaches its maximum in the low-to-intermediate energy region, and then decreases at higher $N$.
For PNR(1) and PNR(2), the admissible range rapidly shrinks at large $N$, whereas PNR(3) retains an additional tolerance window.
A similar trend is observed for the phase mismatch in Fig.~\ref{fig:max_squeezing}(b), but the tolerable phase error collapses more rapidly with increasing $N$, in agreement with the enhanced phase sensitivity predicted by Eq.~\eqref{eq:rm_approx}.
Moreover, the extra tolerance window preserved by PNR(3) is consistent with the parity-step behavior discussed above: increasing the PNR resolution improves mismatch robustness in a non-smooth, threshold-like manner rather than continuously.

Overall, the simulations show that the IS-Kennedy receiver is far more sensitive to phase mismatch than to magnitude mismatch, and that finite-resolution PNR improves robustness mainly against magnitude mismatch rather than imperfect phase locking.

\section{Discussion and Conclusion}
In this study, we have clarified the residual-squeezing mechanism underlying the performance degradation of IS-Kennedy receivers due to squeezing mismatch. By formulating a unified mismatch model, we showed that imperfect inverse squeezing alters the photon-number statistics of the nulled states, necessitating a transition from single-threshold to complex MAP decision rules.

Our research identifies two key physical effects that dictate the receiver's robustness. First, we revealed a ``dominant masking effect": the phase mismatch is scaled by a factor that grows rapidly with signal energy ($\frac{1}{2}\sinh(2r)$), making phase control the primary determinant of performance in the high-energy regime. Unlike magnitude mismatch, which results in a saturable error floor, phase mismatch leads to a polynomial growth of the leading error contribution and can therefore rapidly destroy the SQL advantage. Second, we characterized the ``parity-step effect," showing that PNR detectors enhance robustness in a non-continuous, threshold-like manner. This effect is a direct consequence of the even-photon distribution of the residual squeezed vacuum state.

Together with our previous analysis of channel phase diffusion and transmission loss \cite{IS_Kennedy}, the present results further clarify the practical operating regime of the IS-Kennedy receiver. Its main advantage lies in converting transmitter-side squeezing into an enhanced effective coherent-state separation before photon counting, thereby enabling discrimination performance beyond coherent-state benchmarks under favorable conditions. In the presence of transmission loss, this squeezing-enabled advantage is progressively reduced; nevertheless, the IS-Kennedy can still outperform the conventional Kennedy receiver over a finite propagation distance, particularly in the low-signal-energy regime \cite{IS_Kennedy}. Therefore, rather than targeting ultra-long-haul unamplified links, where coherent-state signaling benefits from intrinsic robustness to pure loss, the IS-Kennedy receiver is better suited to links with limited propagation loss and sufficiently stable phase control.

These considerations also suggest clear directions for further improvement. The strong sensitivity to squeezing-phase mismatch identified here indicates that high-precision or adaptive phase locking and phase compensation should be prioritized over simply increasing the photon-number resolution. Another natural extension is to jointly optimize the Kennedy displacement together with the inverse-squeezing parameters, which may partially compensate for mismatch-induced changes in the photon-number statistics. More generally, joint optimization of the transmitted squeezing resource and receiver parameters under channel loss and squeezing mismatch may further enlarge the operating regime in which the IS-Kennedy receiver provides a practical advantage.

\appendix

\section{Fock-basis Population of Squeezed Vacuum and Displaced Squeezed States}
Let the Squeezed Vacuum state (SVS) and the Displaced Squeezed State (DSS) be denoted, respectively, as:
\begin{equation}
  S(z)\ket{0} \quad \text{and} \quad D(\alpha)S(z)\ket{0},
\end{equation}
where the squeezing parameter is complex, $z = r\mathrm{e}^{j\theta}$ (with $j=\sqrt{-1}$).

The photon number distribution of the Squeezed Vacuum state is non-zero only for even photon numbers:
\begin{equation}
  \label{eq:pn_svs}
  p_{2k}^{\text{svs}}(r) = \frac{(2k)!}{2^{2k}(k!)^2} \frac{\tanh^{2k}(r)}{\cosh(r)},
  \quad p_{2k+1}^{\text{svs}}(r) = 0.
\end{equation}

The photon number distribution of the displaced squeezed state can be expressed in terms of Hermite polynomials, $H_n(\cdot)$:
\begin{equation}
  p_{n}^{\text{dss}} = \frac{1}{n!\cosh r} \left(\frac{\tanh r}{2}\right)^n \left| H_n\left( \frac{\alpha \mathrm{e}^{-j\theta/2}}{\sqrt{\sinh r \cosh r}} \right) \right|^2 \exp\left( -|\alpha|^2 + \mathrm{Re}\left[ \mathrm{e}^{-j\theta}\alpha^2 \right]\tanh r \right).
  \label{eq:pn_dss}
\end{equation}

\begin{acknowledgments}
This work was supported by 
The Key Program of the Joint Funds of the National Natural Science Foundation of China (No.~U25D8016),
Guangxi Science and Technology Program (Grant No.~GuiKeFN2600640534),
Innovative Talent Development Fund of Information Support Force Engineering University (Grant No.~XJKT-QT-25-02-GW, No.~XJKT-QT-25-03-GW),
and Guangxi Science and Technology Base and Talent Project (Grant No.~GuiKeAD25069071).
\end{acknowledgments}

% \section*{Author contributions}
% E. B. conceived the study, performed the theoretical analysis and numerical simulations, and wrote the initial manuscript draft. J. P. and T. W. provided critical guidance on the research direction and methodology. C. D. and Z. Z. supervised the project, interpreted the results, and secured the funding. T. W., J. P., K. W., F. S., C. Z. and Y. L. offered expertise in specialized theoretical aspects and reviewed the manuscript.

\section*{Data Availability}
The data that support the findings of this study are available from the authors upon request.
\section*{Code Availability}
The numerical simulations in this study were performed in Python using QuTiP 5.1.1. The custom scripts used to generate the results are available from the corresponding author on reasonable request.
\section*{Competing interests}
The authors declare no competing interests.

% ​\nocite{*}

\bibliography{refs}% Produces the bibliography via BibTeX.

\end{document}